\pdfoutput=1

\documentclass[11pt]{article}

\usepackage[final]{acl}
\usepackage{listings}
\usepackage{times}
\usepackage{latexsym}
\usepackage{bm}
\usepackage[T1]{fontenc}
\usepackage{natbib}
\usepackage[utf8]{inputenc}

\usepackage{microtype}

\usepackage{inconsolata}

\usepackage{listings}

\usepackage{times}
\usepackage[T1]{fontenc}
\usepackage[utf8]{inputenc}
\usepackage{graphicx}

\usepackage{hyperref}
\usepackage{amsmath}
\usepackage{multirow}
\usepackage{booktabs}
\usepackage{subcaption}

\title{CELI: Controller-Embedded Language Model Interactions}

\author{
  Jan-Samuel Wagner,
  Dave DeCaprio,
  Abishek Chiffon Muthu Raja,
  Jonathan M. Holman,
  Lauren K. Brady, \\
  \textbf{Sky C. Cheung},
  \textbf{Hosein Barzekar},
  \textbf{Eric Yang},
  \textbf{Mark Anthony Martinez II},
  \textbf{David Soong}, \\
  \textbf{Sriram Sridhar},
  \textbf{Han Si},
  \textbf{Brandon W. Higgs},
  \textbf{Hisham Hamadeh}, \\
  \textbf{Scott Ogden} \\
  Genmab US, Inc. \\
  777 Scudders Mill Rd, Plainsboro Township, NJ 08536 \\
\texttt{\{jwag, ddec, abic, joho, lbr, skch, hoba, erya, mkmz, tsoo\}@genmab.com} \\
\texttt{\{srsr, hasi, brhi, hha, scog\}@genmab.com}
}

\begin{document}
\maketitle
\begin{abstract}
We introduce Controller-Embedded Language Model Interactions (CELI), a framework that integrates control logic directly within language model (LM) prompts, facilitating complex, multi-stage task execution. CELI addresses limitations of existing prompt engineering and workflow optimization techniques by embedding control logic directly within the operational context of language models, enabling dynamic adaptation to evolving task requirements. Our framework transfers control from the traditional programming execution environment to the LMs, allowing them to autonomously manage computational workflows while maintaining seamless interaction with external systems and functions. CELI supports arbitrary function calls with variable arguments, bridging the gap between LMs' adaptive reasoning capabilities and conventional software paradigms' structured control mechanisms. To evaluate CELI's versatility and effectiveness, we conducted case studies in two distinct domains: code generation (HumanEval benchmark) and multi-stage content generation (Wikipedia-style articles). The results demonstrate notable performance improvements across a range of domains. CELI achieved a 4.9 percentage point improvement over the best reported score of the baseline GPT-4 model on the HumanEval code generation benchmark. In multi-stage content generation, 94.4\% of CELI-produced Wikipedia-style articles met or exceeded first draft quality when optimally configured, with 44.4\% achieving high quality. These outcomes underscore CELI's potential for optimizing AI-driven workflows across diverse computational domains.
\end{abstract}

\section{Background}

The rapid evolution of language models (LMs) has expanded their capabilities from basic question-answering to complex problem-solving tasks. This progression has driven research towards increasingly sophisticated approaches for leveraging LM capabilities through advanced prompt engineering techniques and orchestration layers. Early applications of LMs primarily relied on manually crafted single prompts, where performance was highly dependent on prompt precision and contextual relevance \citep{brown2020language}. Subsequent research introduced more advanced strategies such as Chain-of-Thought \citep{wei2022chain}, Tree of Thoughts \citep{yao2023tree}, ReAct \citep{yao2023react}, and Reflexion \citep{shinn2023reflexion}, each enhancing LMs' problem-solving and decision-making abilities. As the field matured, structured approaches for prompt management and workflow optimization emerged. Frameworks like DSPy \citep{khattab2023demonstrate} and Trace \citep{cheng2024trace} offered significant improvements in systematization and optimization.

However, existing frameworks frequently encounter challenges in real-time adaptation, especially when managing complex, multi-stage tasks. These limitations include the inability to handle unforeseen scenarios, lack of dynamic task prioritization, and reduced context awareness across multi-stage processes. To address these challenges, we introduce the Controller-Embedded Language Model Interactions (CELI) framework. CELI embeds control logic directly within LM prompts, enabling dynamic execution of complex tasks and seamless interfacing with object-oriented programming environments. This approach allows CELI to overcome the limitations of static prompt structures, offering a flexible, adaptive, yet powerful solution for managing intricate, multi-stage tasks that require both nuanced natural language processing and precise programmatic execution.

\section{Related Work}

Recent work has explored discrete optimization and reinforcement learning for single LM calls. The concept of \emph{foundation model programming} \citep{khattab2023demonstrate} has emerged, treating LMs as integral components of larger computational systems rather than standalone question-answering tools. Existing frameworks like LangChain \citep{chase2022langchain} and Semantic Kernel \citep{microsoft2023semantickernel} have made significant progress in connecting LMs with external tools. These frameworks have improved the efficiency and reliability of LM-driven tasks, expanding their potential applications. The versatility of these systems is often demonstrated through support for various learning paradigms, including zero-shot, one-shot, and multi-shot approaches, building upon the capabilities of in-context learning \citep{mccann2018natural, radford2018improving, brown2020language}. Recent work has also explored discrete optimization and reinforcement learning for single LM calls \citep{guo2023automatic, pryzant2023differentiable}, as well as enhancing LMs' interaction with external knowledge sources \citep{lewis2023retrieval}.

While these approaches aim to enhance LM performance in specific scenarios, they often fall short in addressing the broader limitations of existing frameworks. These limitations include difficulties in maintaining coherence in complex, expansive processes, inability to autonomously refine strategies based on intermediate results, and challenges in error detection and recovery during task execution. CELI addresses these gaps by introducing a comprehensive approach to embedding control logic within LM workflows, enabling more effective handling of complex, multi-stage tasks requiring real-time adaptations.

\begin{figure*}[t]
    \centering
    \includegraphics[width=\textwidth]{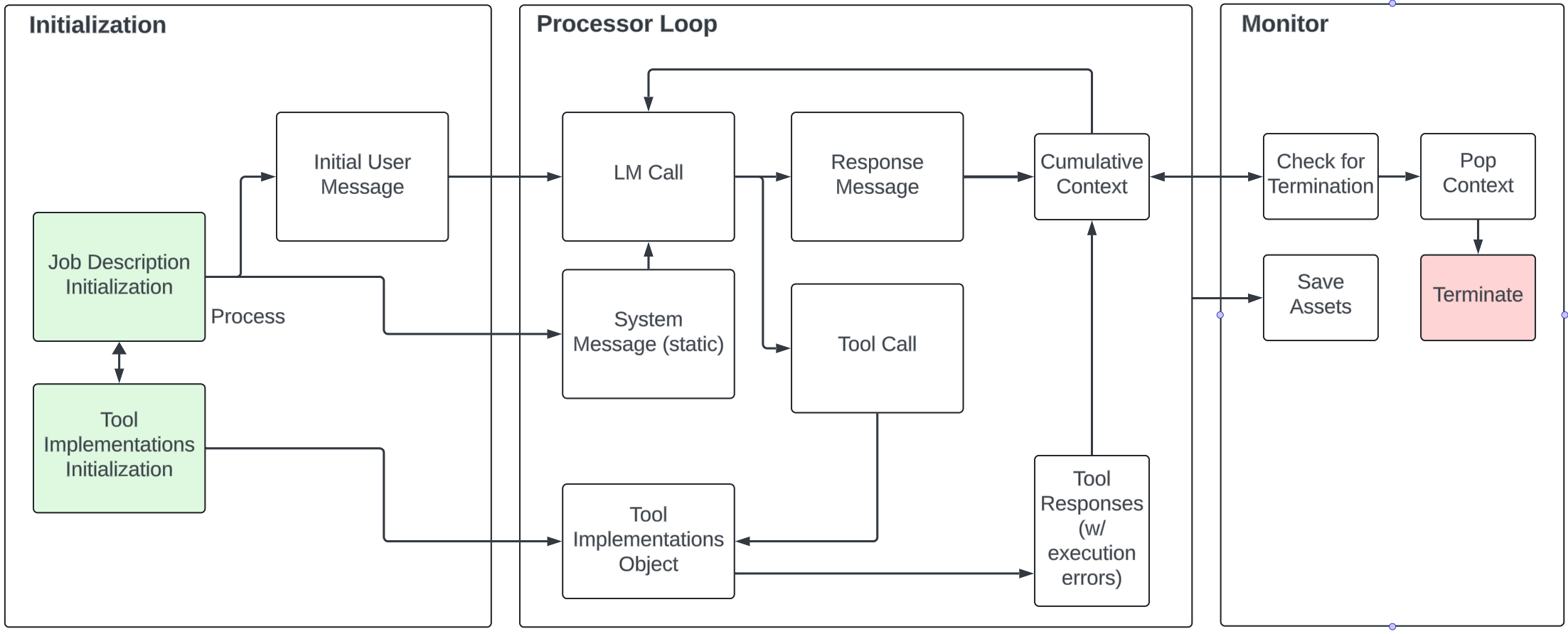}
    \caption{CELI Workflow Architecture}
    \label{fig:workflow}
\end{figure*}

\section{Foundational Elements of the CELI Framework}

CELI addresses several key limitations of existing frameworks through its novel approach to embedding control logic within LM prompts. For instance, while frameworks like LangChain struggle with maintaining coherence across extended multi-stage processes, CELI's cumulative context mechanism ensures consistent state management throughout task execution. This allows CELI to handle complex tasks like multi-section document generation without losing context or coherence. Additionally, where traditional frameworks often rely on predefined workflows, CELI's dynamic task prioritization enables real-time adaptation to unforeseen scenarios. For example, in code generation tasks, CELI can autonomously reprioritize subtasks based on intermediate results, such as shifting focus to error handling when unexpected inputs are encountered. Furthermore, CELI's ability to seamlessly interface with external tools and functions allows it to overcome the limited scope of single LM calls. This is particularly evident in tasks requiring specialized knowledge or computations, where CELI can dynamically invoke relevant tools or APIs as needed, maintaining a fluid workflow that adapts to the task's evolving requirements.

\subsection{Embedded-Controller Logic}

CELI framework's embedded controller logic (which “calls the shots”) is implemented through a \textbf{Job Description} module, which serves as the blueprint for task execution. This Job Description is compiled into two key components: a \textbf{system message} and an \textbf{initial user message}. The system message encapsulates the LM’s role, objectives, task list, and operational rules, while the initial user message initiates the workflow. Together, these elements form the foundation of CELI's control mechanism, guiding the LM's behavior throughout the execution process. For a detailed explanation of these components and their roles in the CELI framework, see Appendix~\ref{subsubsec:job_description_module} and~\ref{subsubsec:master_template_factory_module}.

\subsection{Workflow Architecture}

Figure~\ref{fig:workflow} illustrates the CELI framework's architecture and workflow, starting with the initialization of both the Job Description and Tool Implementations, which define the tasks and available tools for the system. This feeds into the Processor Loop, where the core task execution takes place. During this loop, the LM may invoke specific tools to complete its tasks, interacting with the Tool Implementations Object that stores these functions. Responses are generated based on tool outputs or direct LM reasoning and are continuously fed into the cumulative context, which maintains the state across multiple iterations of the task execution. Errors during tool execution are handled within this context, ensuring smooth progression through the workflow.

The Monitor component oversees the workflow's progress, checking for termination conditions. It monitors the process to ensure tasks are fully completed, popping context when all tasks are complete, preventing context window limits from being reached. Before termination, the system saves relevant assets, such as logs or results. Once all tasks are finalized, the process terminates. This architecture highlights CELI's ability to dynamically handle complex tasks by combining LM-based processing with tool execution, while maintaining context and ensuring robust monitoring for task completion.

A comprehensive overview of the CELI framework architecture and its core components is provided in Appendix~\ref{subsec:framework_architecture} and~\ref{subsec:core_components}.

\subsection{Iterative Task Execution}

\begin{figure}[t]
    \centering
    \includegraphics[width=\linewidth]{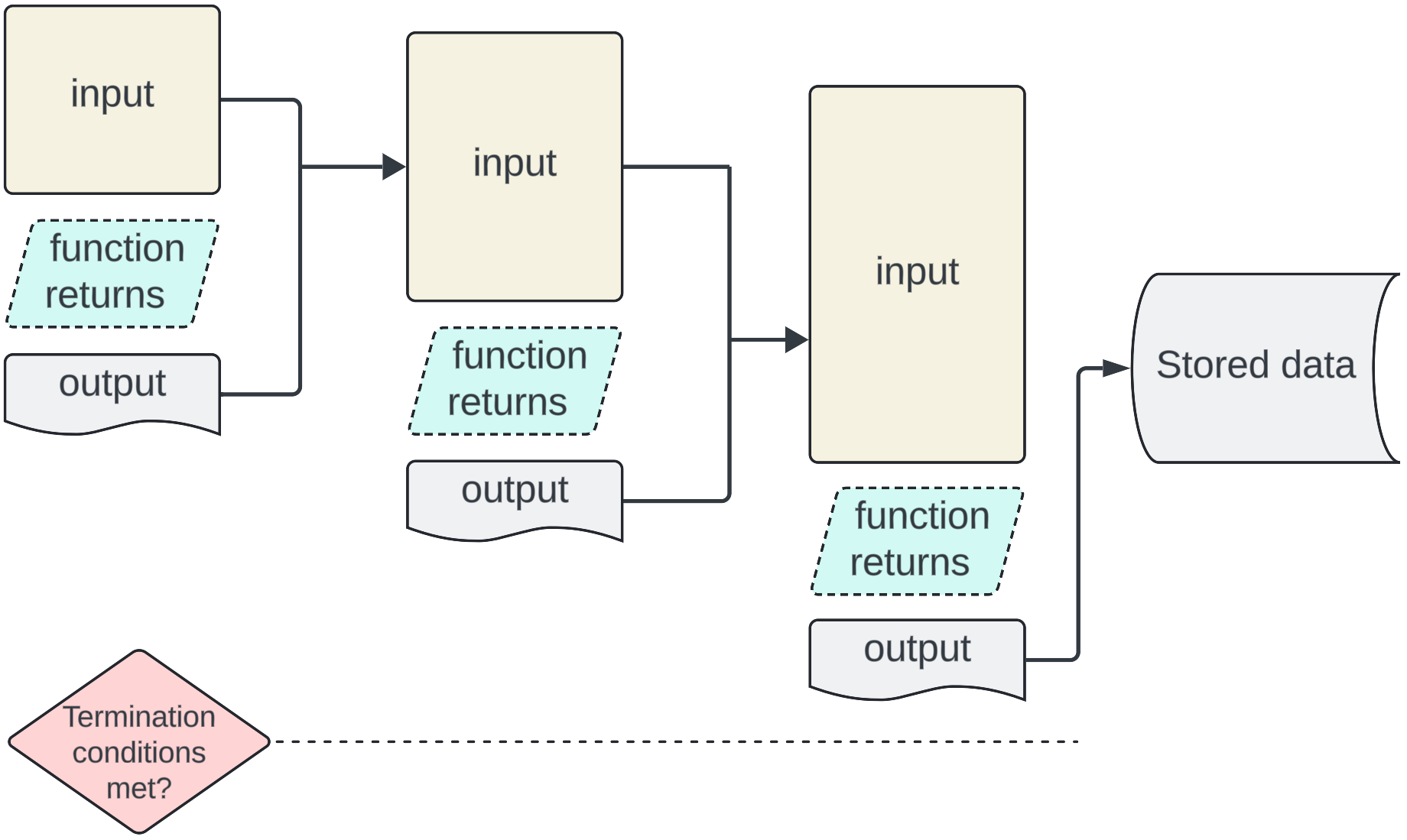}
    \caption{Iterative Task Execution}
    \label{fig:task_exe}
\end{figure}

Figure~\ref{fig:task_exe} illustrates the detailed iterative task execution process within the CELI framework. Each task begins with an initial input, which is processed either directly or through a function call, as needed. The function returns, represented with a dotted outline, indicate that the function execution is optional and may not always occur, depending on the task requirements. The output is generated from this process and is either passed forward to the next iteration or further refined. Throughout the execution, CELI performs a continuous check for termination conditions to determine if the task has reached completion. If the termination criteria are met, the process concludes, and the results are stored. Otherwise, the task continues to iterate with new inputs and optional function executions until it reaches the termination point. This figure highlights the flexibility of CELI in dynamically adapting its execution based on task needs, ensuring efficient task completion while maintaining adaptability for complex workflows. For an in-depth explanation of CELI's workflow execution process, refer to Appendix~\ref{subsec:workflow_execution}.

\subsection{Addressing Limitations and Fostering Emergent Capabilities}

This architecture and workflow directly address many of the limitations observed in traditional LM frameworks. By embedding control logic within prompts and employing an iterative execution process, CELI enables real-time adaptations to unforeseen scenarios and evolving task requirements. The Processor Loop and Monitor component work in tandem to maintain context awareness across multi-stage tasks, facilitating coherent execution of complex, long-running processes. The framework's ability to dynamically invoke tools and refine strategies based on intermediate results supports autonomous learning and improvement during task execution. Furthermore, CELI's flexible architecture fosters emergent properties such as context-awareness, self-guided learning, and adaptability. These features collectively enable CELI to handle intricate, multi-stage tasks with a level of sophistication and efficiency that surpasses traditional prompt-based approaches, paving the way for more advanced AI-driven problem-solving across diverse domains. The advanced features of CELI, including monitoring, embedding, and mapping capabilities, are further detailed in Appendix~\ref{subsec:advanced_features}.

\section{Case Studies}

\subsection{Native Integration with Traditional Programming Execution Environment (HumanEval)}

\subsubsection{Background and Objectives}

The HumanEval dataset \citep{chen2021evaluating}, introduced by OpenAI, provides a benchmark for evaluating the coding abilities of LMs through 164 Python coding challenges. These challenges span a wide range of programming concepts and algorithms, offering a comprehensive test of code generation capabilities. While primarily designed for code assessment, HumanEval presents an opportunity to showcase CELI's broader potential in integrating LMs with diverse programming tools and paradigms.

Our objectives for this case study were to demonstrate CELI's effectiveness in tackling coding challenges through its native integration with execution environments, evaluate its performance against traditional LM approaches and state-of-the-art frameworks, and assess its ability to implement an iterative process of solution generation, testing, and refinement.

\subsubsection{Implementation}

The implementation is built around a JobDescription module defining key tasks and HumanEvalTools class providing essential functionality. The framework retrieves coding prompts, develops test cases, iteratively writes and tests code, and saves final outputs. It employs sandboxed execution for safety and supports both user-defined and official tests. This implementation showcases CELI's ability to integrate effectively with LMs and programming environments while maintaining security and flexibility in automated code generation and testing. JobDescription tasks and HumanEvalTools functions are highlighted in Appendix~\ref{subsection:humaneval_job_description} and \ref{subsection:humnaneval_tool_implementations}.

\subsubsection{Evaluation Process}

CELI's performance on the HumanEval benchmark was evaluated through two main approaches. First, we performed a qualitative analysis of the logs generated during CELI's problem-solving process. These logs offered insights into how CELI interpreted problems, generated test cases, implemented solutions, and refined code based on test results, illustrating its reasoning and adaptation capabilities (\emph{Log Analysis}). Second, we conducted a quantitative assessment by running CELI's solutions against the official HumanEval test cases and calculating the number of correct solutions. This provided a clear measure of CELI's ability to generate functionally correct code across diverse programming challenges (\emph{Quantitative Analysis}).

To contextualize CELI's performance, we compared it against baseline GPT-4 models and other state-of-the-art frameworks. We used both GPT-4-Turbo (1106) and (0125) as baselines, with scores obtained from a comprehensive benchmark study \citep{OpenAISimpleEvals}. This dual-baseline approach allowed for fair comparisons, as existing frameworks primarily used the 1106 version, while we found the 0125 version more suitable for our specific implementation (further discussed in Appendix~\ref{subsection:Selection_of_Evaluation}). By examining CELI's results in relation to these baselines and other frameworks, we gained a comprehensive understanding of its performance in the context of current LM capabilities.

\subsubsection{Results}

\paragraph{Log Analysis}

Analysis of the logs revealed insights into CELI's problem-solving process across various HumanEval tasks. The logs demonstrated CELI's ability to interpret problems, generate test cases, implement solutions, and refine code based on test results. This qualitative analysis showcased CELI's reasoning and adaptation capabilities, as illustrated in 


A representative example from the logs for task \texttt{HumanEval/2} is paraphrased below (see Appendix~\ref{subsection:example_log} for the full log):

\begin{enumerate}
    \item CELI analyzed the problem and developed test cases.
    \item Identified key scenarios to test various number types.
    \item Created \texttt{check(candidate)} function with appropriate test cases.
    \item Implemented function and attempted to run tests; encountered error.
    \item Corrected error, ran tests again; identified one test case failure.
    \item Analyzed failure, attributed to floating-point issues; updated function accordingly.
    \item Updated function passed all tests; CELI saved final implementation.
\end{enumerate}

This log analysis reveals several key strengths of CELI's approach to coding challenges. Firstly, CELI demonstrates proactive error prevention by developing comprehensive test cases (steps 1-3) that cover various number types, showcasing its ability to anticipate edge cases. Secondly, CELI's error handling and debugging capabilities are evident in its response to the initial test failure (steps 4-5). Rather than simply regenerating code, CELI analyzed the failure, identified the specific issue with floating-point numbers, and made targeted corrections. This targeted refinement process likely contributes to CELI's high success rate on the HumanEval benchmark, as it allows for efficient and effective code improvement without unnecessary rewrites. Furthermore, CELI's ability to maintain context throughout the problem-solving process (steps 1-7) enables it to build upon its initial approach rather than starting from scratch with each iteration. This contextual awareness and iterative refinement capability align with CELI's strong performance relative to specialized coding frameworks, suggesting that these problem-solving strategies are key factors in its effectiveness across diverse programming challenges.

\paragraph{Quantitative Analysis}

CELI's performance on the HumanEval benchmark was evaluated against baseline GPT-4 models and state-of-the-art coding frameworks. Table~\ref{tab:humaneval_results} presents the results of this comparison.

\begin{table}[t]
\centering
\scalebox{.85}{
\begin{tabular}{lcc}
\toprule
\textbf{Approach} & \textbf{Pass@1} & \textbf{$\Delta$ $\boldsymbol{\nabla}$} \\
\midrule
GPT-4-Turbo (1106) Base* & 83.7\% & -- \\
GPT-4-Turbo (0125) Base** & 86.6\% & -- \\
MapCoder* \citep{islam2024mapcoder} & 93.9\% & +10.2 \\
AgentCoder* \citep{huang2024agentcoder} & 89.6\% & +5.9 \\
LDB* \citep{zhong2024ldb} & 89.6\% & +5.9 \\
AgentVerse* \citep{chen2023agentverse} & 89.0\% & +5.3 \\
\textbf{CELI**} & \textbf{91.5\%} & \textbf{+4.9} \\
\bottomrule
\end{tabular}
}
\caption{CELI vs. Baseline GPT-4 Turbo LM HumanEval Pass@1 scores. Baseline scores for GPT-4-Turbo models obtained from OpenAI's simple-evals repository (OpenAI, 2024b). *Comparison with GPT-4-Turbo (1106). **Comparison with GPT-4-Turbo (0125).}
\label{tab:humaneval_results}
\end{table}

CELI outperformed the best reported score of the baseline GPT-4-Turbo (0125) model by 4.9 percentage points on the HumanEval benchmark (Pass@1). Notably, CELI's performance improvement (delta score) was within 1 percentage point of 3 of 4 comparison frameworks, highlighting its strong competitiveness despite its broader focus. Key factors contributing to CELI's strong performance include its ability to generate comprehensive test cases, implement iterative solution refinement based on test results, and maintain context across multiple iterations.

These capabilities allow CELI to approach coding tasks with sophistication comparable to specialized frameworks while retaining flexibility for diverse task types. The results demonstrate CELI's potential as a versatile framework for complex, multi-stage tasks requiring both natural language understanding and precise programmatic execution. For transparency and reproducibility, the complete code, result set, and cached run have been made available in the CELI GitHub repository \citep{x3n0cr4735_celi2024}. Interested parties can run the implementation using cached results to observe the exact output of the process task-by-task.

\subsection{Document Generation with Multi-Stage Processing (Wikipedia Articles)}

\subsubsection{Background and Objectives}

Wikipedia articles present unique challenges for language models due to their demand for factual accuracy, consistent structure, and comprehensive coverage across diverse topics. These articles require synthesizing information from multiple sources, maintaining a neutral point of view, and adhering to specific formatting guidelines. Wikipedia's breadth of subjects also tests a system's adaptability to various domains.
We chose Wikipedia as an exemplar for evaluating CELI's capabilities in multi-stage document processing because it represents a gold standard for structured, factual content creation. Our objective was to assess CELI's performance in generating high-quality articles across diverse topics (Bands, Drugs, and Countries), using existing articles as templates. This study aimed to demonstrate CELI's effectiveness in maintaining coherence, accuracy, and appropriate structure in complex, multi-domain content generation tasks.

\subsubsection{Implementation}

The implementation is built around a JobDescription defining key tasks and WikipediaTools class providing essential functionality. The framework retrieves example and target Wikipedia pages, analyzes section structures, develops content, and compiles final articles. It employs content retrieval, question-asking, and drafting mechanisms for comprehensive article generation.
This implementation showcases CELI's ability to integrate effectively with LMs for complex document generation while maintaining flexibility in automated content creation and structuring. The use of two advanced LMs, GPT-4 \citep{openai2024gpt4} and Claude 3.5 Sonnet \citep{anthropic2024claude}, allows CELI to benefit from their complementary strengths in natural language understanding and generation. Detailed descriptions of JobDescription tasks and WikipediaTools functions are provided in Appendix~\ref{subsection:job_description} and \ref{subsection:tool_implementations}.

\subsubsection{Evaluation Process}

CELI's performance in generating Wikipedia-style articles was evaluated through two complementary approaches. First, we conducted a qualitative analysis by examining the logs generated during CELI's article creation process, providing insights into its content structuring, information gathering, and adaptation strategies (\emph{Log Analysis}). Second, we performed a quantitative assessment of 18 articles across three categories (Bands, Drugs, and Countries) using GPT-4 and Claude 3.5 Sonnet on a 0-6 scale (\emph{Quantitative Analysis}). Scores of 4 and above were considered to meet or exceed first draft quality, with 4 defined as "Satisfactory: Acceptable as a first draft but requires refinement." Scores of 5 and above were classified as achieving high quality, indicating comparable or superior quality to the ground truth. We analyzed performance across categories and models, including a best-case scenario selecting the highest-performing model for each article. This dual approach allowed us to assess both CELI's problem-solving process and the quality of its output. Full evaluation criteria, including limitations and potential for hallucination, are detailed in Appendix~\ref{subsection:Evaluation_Prompt}.

\subsubsection{Results}

Our evaluation of CELI's performance in generating Wikipedia-style articles combined qualitative log analysis with quantitative assessment.

\paragraph{Log Analysis}

Analysis of the logs revealed CELI's sophisticated approach to multi-stage document generation, particularly in handling complex, long-form content like Wikipedia articles. The logs demonstrated CELI's ability to structure content, identify information gaps, ask relevant questions, and adapt to challenges. 
A representative example from the logs for the \emph{Jonas Brothers} article generation (with Led Zeppelin as example article) is paraphrased below (see Appendix~\ref{subsection:Example_Log} for the full log):

\begin{enumerate}
    \item CELI analyzed the example document structure and content.
    \item Identified key sections to include in the target article.
    \item Generated relevant questions (to query vector database) in order to fill information gaps.
    \item Encountered errors in retrieving additional information.
    \item Adapted approach to use available information from references and example document.
    \item Defined subsections for the article based on available information.
    \item Drafted and saved the article section, demonstrating ability to complete task despite challenges.
\end{enumerate}

\begin{table*}[t]
\centering
\scalebox{1.1}{
\begin{tabular}{lccc}
\toprule
\textbf{Category} & \textbf{GPT-4} & \textbf{Sonnet 3.5} & \textbf{Best Model} \\
\midrule
Bands & 33.3\% / 0.0\% & 83.3\% / 16.7\% & 83.3\% / 16.7\% \\
Countries & 100.0\% / 66.7\% & 50.0\% / 16.7\% & 100.0\% / 66.7\% \\
Drugs & 60.0\% / 20.0\% & 100.0\% / 50.0\% & 100.0\% / 50.0\% \\
\midrule
Overall & 64.7\% / 29.4\% & 77.8\% / 27.8\% & 94.4\% / 44.4\% \\
\bottomrule
\end{tabular}
}
\caption{Percentage of Articles Meeting First Draft Quality ($\geq$ 4) and Achieving High Quality ($\geq$ 5). Each cell shows "\% meeting or exceeding first draft quality / \% achieving high quality."}
\label{tab:article_quality}
\end{table*}

Analysis of the logs revealed CELI's sophisticated approach to multi-stage document generation, particularly in handling complex, long-form content like Wikipedia articles. The representative example from the \emph{Jonas Brothers} article generation process illustrates several key strengths: CELI's ability to analyze and adapt to complex document structures, proactively gather information by generating relevant questions, and resiliently handle information retrieval challenges. Notably, CELI demonstrated context-aware content structuring by defining subsections based on available information, and showed persistence in completing the task despite obstacles. These capabilities suggest CELI's potential for producing well-organized, comprehensive articles across diverse topics. The framework's approach of combining structural understanding, proactive information gathering, and adaptive refinement not only showcases its effectiveness in Wikipedia article generation but also indicates its versatility for tackling a wide range of complex document creation tasks. This qualitative analysis provides valuable insights into CELI's problem-solving strategies and adaptability, setting the stage for a deeper understanding of its performance in generating Wikipedia-style articles.

\paragraph{Quantitative Analysis}

The evaluation of CELI-generated Wikipedia-style articles across three categories (Bands, Drugs, and Countries) revealed promising results. Table~\ref{tab:article_quality} presents the percentage of articles meeting or exceeding first draft quality (score $\geq$ 4) and achieving high quality (score $\geq$ 5) for each category and model.

Performance varied notably across categories and models. GPT-4 excelled in the Countries category, with all articles meeting or exceeding first draft quality and 66.7\% achieving high quality. Claude 3.5 Sonnet performed exceptionally well in the Drugs category, with all articles meeting or exceeding first draft quality and 50\% achieving high quality. Both models faced challenges in the Bands category, though Claude 3.5 Sonnet showed more consistent performance. The Best Model approach, which selects the highest-performing model for each category, demonstrates CELI's potential when leveraging the strengths of different models. This approach resulted in 94.4\% of articles meeting or exceeding first draft quality, with 44.4\% achieving high quality across all categories.

Analysis of Claude 3.5 Sonnet's evaluation rationales provided valuable insights into the strengths and weaknesses of CELI-generated articles. Higher-scoring articles demonstrated well-organized structures, broad content coverage, and strong factual accuracy, particularly in specialized categories like Drugs. For instance, a \emph{Prednisone} article was noted for its "thorough and structured approach" to discussing the drug's various aspects. Lower-scoring articles, especially in the Bands category, often suffered from factual errors and structural inconsistencies. A \emph{Daft Punk} article, for example, contained "several inaccuracies and inconsistencies" compared to the reference document. In the Countries category, while many articles were highly rated, a particularly low-scoring article (0) was generated when the model failed to maintain relevance, producing content about an entirely different country than the one requested. Full text of Claude 3.5 Sonnet's evaluation rationales, for three example articles, is available in Appendix~\ref{subsection:Example_Evaluations}, providing deeper insights into the quality assessment process.

\section{Discussion}

The CELI framework represents a significant advancement in integrating LMs into complex computational systems. By embedding control logic directly within LM prompts, CELI addresses key limitations identified in existing frameworks such as LangChain \citep{chase2022langchain}, Semantic Kernel \citep{microsoft2023semantickernel}, DSPy \citep{khattab2023demonstrate}, and Trace \citep{cheng2024trace}, particularly in handling complex, multi-stage tasks with real-time adaptations.

The application of CELI in our case studies illustrates its effectiveness across three distinct areas: integrating control mechanisms within language model prompts, facilitating iterative problem-solving through interactive recursion, and managing the execution of complex, multi-phase tasks. In the HumanEval benchmark, CELI demonstrated the ability to autonomously navigate through the steps of code generation, testing, and correction, achieving a notable improvement over the baseline GPT-4 model. The embedded control logic enables CELI to adjust its behavior in real-time, handling exceptions and reordering tasks based on intermediate outcomes. In the task of multi-stage document generation, CELI exhibited strong performance by coordinating content development in a structured manner, producing high-quality outputs in the majority of cases. These findings suggest that CELI is well-suited for optimizing workflows that require both adaptive reasoning and precise task management across varied computational domains.

Several emergent properties of CELI were observed across the case studies. CELI demonstrated strong context-awareness in the HumanEval benchmark, maintaining contextual understanding throughout problem analysis, test case creation, function implementation, and result-based refinement. Self-guided in-context learning was evident in the Wikipedia article generation task, where CELI autonomously adapted its approach when faced with information retrieval challenges, leveraging available references and examples. Furthermore, CELI exhibited adaptability and dynamism in both case studies, showcasing its ability to adjust strategies and logic in response to new patterns or challenges encountered during task execution.

These properties underscore a fundamental shift in how we conceptualize and utilize LMs. The transition from traditional "prompts" to "Language Model Interactions" in CELI represents a new paradigm that allows for more sophisticated, context-aware, and goal-oriented engagements with LMs. By enabling LMs to autonomously manage complex workflows while maintaining seamless interaction with external systems and functions, CELI opens up new possibilities for AI-driven task automation and problem-solving across diverse computational domains.

Building on these advancements, future work should focus on further enhancing CELI's capabilities. This includes improving its robustness in handling highly dynamic task environments, refining its error detection and recovery mechanisms, and exploring its potential in multi-modal tasks. Additionally, investigating CELI's applicability in collaborative human-AI settings could provide valuable insights into its real-world utility and potentially expand its impact across various fields and industries.

\section{Limitations}

While CELI shows promising results, several areas warrant further investigation. First, although our case studies span diverse domains (coding and content generation), the framework's performance across an even broader range of task types requires evaluation to fully assess its generalizability. Ongoing work is focused on extending CELI's application to additional domains, such as decision-making tasks and large-scale data processing, with preliminary results indicating promising adaptability to varied task structures.

Second, embedding control logic within prompts may introduce computational overhead, particularly in large-scale applications. This trade-off between enhanced functionality and increased computational demands is being addressed through continuous optimization efforts. These include refining prompt engineering techniques, exploring task decomposition methods, and improving language model efficiency to minimize overhead while maintaining CELI's advanced capabilities.

Third, the potential for error propagation in complex, multi-stage tasks requires further analysis. While CELI has demonstrated the ability to autonomously handle and adapt to errors, minor errors accumulating across multiple stages could lead to inaccuracies in final outputs. To mitigate this, we are developing more robust error detection and correction mechanisms, such as real-time feedback loops and enhanced monitoring, to ensure reliable outputs in complex workflows.

Fourth, CELI's current implementation is constrained by the context window limits of the underlying LMs, which may hinder its performance in tasks requiring extensive cumulative context. To address this, we are actively investigating techniques such as recursive summarization and hierarchical task decomposition, which will allow CELI to handle larger volumes of data and extended task durations more effectively.

Lastly, CELI's performance is closely linked to the capabilities of the underlying LMs. As demonstrated in our experiments, the choice of model can significantly impact outcomes. However, CELI’s modular architecture allows for rapid integration of new models, and regular benchmarking is conducted to ensure optimal performance as new LMs are released. In addition, CELI's ability to dynamically generate and execute code raises important ethical and safety considerations. To mitigate these risks, we have implemented sandboxed execution environments, and utilize human oversight, to ensure the safe and accurate execution of tasks.

Additionally, while not unique to CELI, the framework's ability to generate and execute code dynamically raises important ethical and safety considerations. The potential for generating harmful or malicious code, albeit unintentionally, necessitates robust safety measures and careful monitoring. In applications involving content generation, such as the Wikipedia article creation case study, ensuring factual accuracy and avoiding biases becomes crucial.

Addressing these limitations is essential to unlocking the full potential of CELI and similar frameworks in AI-driven task automation and problem-solving.

It should be noted that, in the preparation of this paper, we utilized GPT-based AI tools to assist with language refinement. These tools provided suggestions for improving clarity, grammar, and phrasing, as well as offering edits for better readability. All AI-generated suggestions were carefully reviewed by the authors to ensure they aligned with the content and research objectives. No AI-generated content was accepted without thorough validation by the authors.

\section*{Acknowledgments}
We thank the entire team at Genmab US, Inc. for their support and contributions to this research.


\bibliography{references}

\begin{thebibliography}{23}
\providecommand{\natexlab}[1]{#1}

\bibitem[{Anthropic(2024)}]{anthropic2024claude}
Anthropic. 2024.
\newblock Claude 3.5 sonnet model card addendum.
\newblock Technical report.
\newblock Retrieved from \url{https://www-cdn.anthropic.com/fed9cc193a14b84131812372d8d5857f8f304c52/Model_Card_Claude_3_Addendum.pdf}.

\bibitem[{Brown(2020)}]{brown2020language}
Tom~B Brown. 2020.
\newblock Language models are few-shot learners.
\newblock \emph{arXiv preprint arXiv:2005.14165}.

\bibitem[{Chase(2022)}]{chase2022langchain}
Harrison Chase. 2022.
\newblock Langchain: Building applications with lms through composability.
\newblock Technical report.
\newblock GitHub repository. Retrieved from \url{https://github.com/hwchase17/langchain}.

\bibitem[{Chen et~al.(2021)Chen, Tworek, Jun, Yuan, Pinto, Kaplan, Zaremba et~al.}]{chen2021evaluating}
Mark Chen, Jared Tworek, Heewoo Jun, Qiming Yuan, Henrique P. D.~O. Pinto, Jared Kaplan, Wojciech Zaremba, et~al. 2021.
\newblock Evaluating large language models trained on code.
\newblock ArXiv preprint arXiv:2107.03374.

\bibitem[{Chen et~al.(2023)Chen, Su, Zuo, Yang, Yuan, Chan et~al.}]{chen2023agentverse}
Weiyang Chen, Yuying Su, Jieru Zuo, Chen Yang, Chong Yuan, Chun-Ming Chan, et~al. 2023.
\newblock Agentverse: Facilitating multi-agent collaboration and exploring emergent behaviors.
\newblock ArXiv preprint arXiv:2308.10848.

\bibitem[{Cheng et~al.(2024)Cheng, Nie, and Swaminathan}]{cheng2024trace}
Cheng~Allen Cheng, Allen Nie, and Arun Swaminathan. 2024.
\newblock Trace is the new autodiff — unlocking efficient optimization of computational workflows.
\newblock Technical report.
\newblock Preprint under review. Retrieved from \url{https://microsoft.github.io/Trace/}.

\bibitem[{Guo et~al.(2023)Guo, Liu, Tan, Niu, Ren, Wang, and Zheng}]{guo2023automatic}
Haonan Guo, Xinchao Liu, Jian Tan, Chao Niu, Shichao Ren, Shu Wang, and Lirong Zheng. 2023.
\newblock Automatic prompt optimization with gradient descent and beam search.
\newblock ArXiv preprint arXiv:2305.03495.

\bibitem[{Huang et~al.(2024)Huang, Zhang, Luck, Bu, Qing, and Cui}]{huang2024agentcoder}
Di~Huang, Jin~Ma Zhang, Michael Luck, Qing Bu, Yuhong Qing, and Huanyu Cui. 2024.
\newblock Agentcoder: Multi-agent-based code generation with iterative testing and optimisation.
\newblock ArXiv preprint arXiv:2312.13010.

\bibitem[{Islam et~al.(2024)Islam, Ali, and Parvez}]{islam2024mapcoder}
Md~Ashraful Islam, Mohammed~Eunus Ali, and Md~Rizwan Parvez. 2024.
\newblock Mapcoder: Multi-agent code generation for competitive problem solving.
\newblock ArXiv preprint arXiv:2405.11403.

\bibitem[{Khattab et~al.(2023)Khattab, Santhanam, Li, Hall, Liang, Potts, and Zaharia}]{khattab2023demonstrate}
Omar Khattab, Keshav Santhanam, Xin Li, David Hall, Percy Liang, Christopher Potts, and Matei Zaharia. 2023.
\newblock Demonstrate-search-predict: Composing retrieval and language models for knowledge-intensive nlp.
\newblock ArXiv preprint arXiv:2212.14024.

\bibitem[{Lewis et~al.(2023)Lewis, Perez, Piktus et~al.}]{lewis2023retrieval}
Patrick Lewis, Ethan Perez, Aleksandra Piktus, et~al. 2023.
\newblock \href {https://doi.org/10.18653/v1/2023.emnlp-main.75} {Retrieval-augmented generation for knowledge-intensive nlp tasks}.
\newblock In \emph{Proceedings of the 2023 Conference on Empirical Methods in Natural Language Processing (EMNLP)}, pages 1217--1233.

\bibitem[{McCann et~al.(2018)McCann, Keskar, Xiong, and Socher}]{mccann2018natural}
Bryan McCann, Nitish~Shirish Keskar, Caiming Xiong, and Richard Socher. 2018.
\newblock The natural language decathlon: Multitask learning as question answering.
\newblock ArXiv preprint arXiv:1806.08730.

\bibitem[{Microsoft(2023)}]{microsoft2023semantickernel}
Microsoft. 2023.
\newblock Semantic kernel.
\newblock Technical report.
\newblock GitHub repository. Retrieved from \url{https://github.com/microsoft/semantic-kernel}.

\bibitem[{OpenAI(2024)}]{OpenAISimpleEvals}
OpenAI. 2024.
\newblock Simple evals.
\newblock \url{https://github.com/openai/simple-evals}.
\newblock Accessed: October 14, 2024.

\bibitem[{OpenAI et~al.(2024)OpenAI, Achiam, Adler et~al.}]{openai2024gpt4}
OpenAI, Joshua Achiam, Sophie Adler, et~al. 2024.
\newblock \href {https://doi.org/10.48550/arXiv.2303.08774} {Gpt-4 technical report}.
\newblock Technical report.
\newblock ArXiv preprint arXiv:2303.08774.

\bibitem[{Pryzant et~al.(2023)Pryzant, Ermon, and Liang}]{pryzant2023differentiable}
Reid Pryzant, Stefano Ermon, and Percy Liang. 2023.
\newblock Differentiable prompt programming.
\newblock ArXiv preprint arXiv:2305.04010.

\bibitem[{Radford et~al.(2018)Radford, Narasimhan, Salimans, and Sutskever}]{radford2018improving}
Alec Radford, Karthik Narasimhan, Tim Salimans, and Ilya Sutskever. 2018.
\newblock Improving language understanding by generative pre-training.
\newblock Technical report.

\bibitem[{Shinn et~al.(2023)Shinn, Cassano, Berman et~al.}]{shinn2023reflexion}
Nicholas Shinn, Fabio Cassano, Elijah Berman, et~al. 2023.
\newblock Reflexion: Language agents with verbal reinforcement learning.
\newblock ArXiv preprint arXiv:2303.11366v4.

\bibitem[{Wei et~al.(2022)Wei, Wang, Schuurmans et~al.}]{wei2022chain}
Jason Wei, Xuezhi Wang, Dale Schuurmans, et~al. 2022.
\newblock Chain-of-thought prompting elicits reasoning in large language models.
\newblock \emph{NeurIPS}.
\newblock Retrieved from \url{https://arxiv.org/abs/2201.11903}.

\bibitem[{x3n0cr4735(2024)}]{x3n0cr4735_celi2024}
x3n0cr4735. 2024.
\newblock Celi: A framework for controller-embedded language (model) interactions.
\newblock Technical report.
\newblock GitHub repository. Retrieved from \url{https://github.com/x3n0cr4735/celi}.

\bibitem[{Yao et~al.(2023{\natexlab{a}})Yao, Yu, Zhao et~al.}]{yao2023tree}
Shunyu Yao, Dian Yu, Jeffrey Zhao, et~al. 2023{\natexlab{a}}.
\newblock Tree of thoughts: Deliberate problem solving with large language models.
\newblock ArXiv preprint arXiv:2305.10601.

\bibitem[{Yao et~al.(2023{\natexlab{b}})Yao, Zhao, Yu et~al.}]{yao2023react}
Shunyu Yao, Jeffrey Zhao, Dian Yu, et~al. 2023{\natexlab{b}}.
\newblock React: Synergizing reasoning and acting in language models.
\newblock Retrieved from \url{https://arxiv.org/abs/2210.03629}.

\bibitem[{Zhong et~al.(2024)Zhong, Wang, and Shang}]{zhong2024ldb}
Liang Zhong, Zeyu Wang, and Jiaqi Shang. 2024.
\newblock Ldb: A large language model debugger via verifying runtime execution step-by-step.
\newblock ArXiv preprint arXiv:2402.16906.

\end{thebibliography}
\onecolumn

\appendix
\section{CELI Framework Details}
\label{sec:appendix_celi_framework}

\subsection{Framework Architecture}
\label{subsec:framework_architecture}

CELI's architecture is predicated on principles of modularity, flexibility, and scalability, enabling it to address a wide array of applications including data analysis, report generation, content creation, and automated evaluations. CELI decomposes complex workflows into manageable modules that interact seamlessly with LMs. This modular approach facilitates precise task execution while maintaining adaptability to dynamic requirements. By defining clear task descriptions and enabling programmatic interactions with LMs and external systems, CELI ensures high-fidelity task completion and operational efficiency. The framework's design supports facile customization and extension, broadening its applicability across diverse domains. CELI's capacity to handle varied tasks stems from its innovative approach to embedding control logic within LM prompts, effectively transforming these models into self-managing modules capable of executing complex, multi-stage workflows.

CELI's operation consists of two main phases: initialization and execution. These phases ensure robust operation and effective task management across a wide range of applications.

\subsubsection{Initialization}
\begin{enumerate}
    \item System Message Creation: The \texttt{MasterTemplateFactory} creates a comprehensive system message using components from the \texttt{JobDescription}, including role, context, and task library.
    \item Tool Preparation: The system generates tool descriptions from the \texttt{ToolImplementations} class, which are later used by the Section Processors.
    \item ProcessRunner Initialization: The \texttt{ProcessRunner} is created with the \texttt{MasterTemplate}, \texttt{ToolImplementations}, and other configuration parameters.
    \item Section Processor Initialization: The \texttt{ProcessRunner} creates \texttt{SectionProcessor} instances for each section defined in the job description.
\end{enumerate}

\subsubsection{Execution}
\begin{enumerate}
    \item Parallel or Sequential Processing: The \texttt{ProcessRunner} executes all \texttt{SectionProcessors} either concurrently (using asyncio tasks) or sequentially, based on configuration.
    \item LLM-Driven Section Execution: Each \texttt{SectionProcessor} interacts with the LM, which controls the execution flow for its specific section.
    \item Dynamic Tool Utilization: \texttt{SectionProcessors} make tool calls as directed by the LM. These calls are processed through the \texttt{ToolImplementations} provided to the \texttt{ProcessRunner}.
    \item Error Handling and Retries: The system includes built-in review mechanisms and supports retrying sections if necessary.
    \item Completion Mechanism: Each section's process continues until either a loop is detected, or the LM invokes the \texttt{complete\_section} tool.
\end{enumerate}

This streamlined process allows CELI to adapt to various applications by modifying the job description and tool implementations while maintaining a consistent overall structure and flow. The framework's ability to handle complex, multi-stage tasks with dynamic requirements sets it apart from more rigid approaches.

\subsection{Core Components}
\label{subsec:core_components}

\subsubsection{Job Description Module}
\label{subsubsec:job_description_module}

The Job Description module is a critical component of the CELI framework, specifying the configuration for task execution across diverse applications. It defines the roles, contexts, and detailed task lists that guide various automated processes.

\textbf{Key Responsibilities:}
\begin{itemize}
    \item Defining roles and contexts for the task at hand.
    \item Specifying a comprehensive list of tasks to be performed.
    \item Managing configuration settings affecting task execution.
\end{itemize}

\textbf{Core Components:}
\begin{itemize}
    \item Role and context definitions.
    \item Task library (List of Task objects).
    \item Tool implementations class specification.
    \item Pre- and post-context instructions.
\end{itemize}

The Job Description module interacts closely with the Processor module, providing it with the necessary information to execute tasks effectively and in the correct order, whether it's solving math problems, answering questions, or performing code generation.

\subsubsection{Master Template Factory Module}
\label{subsubsec:master_template_factory_module}

The MasterTemplateFactory module transforms the Job Description into a system message. This system message serves as a persistent set of instructions for LMs throughout the execution process.

\textbf{Key Responsibilities:}
\begin{itemize}
    \item Converts the Job Description into a single, comprehensive system message that is used repeatedly to instruct LMs and maintain state across multiple interactions.
\end{itemize}

\textbf{Key Components of the Generated System Message:}
\begin{itemize}
    \item Algorithm setup section based on the role and context from the Job Description.
    \item Numbered task list with resolved cross-references.
    \item General comments and additional instructions as specified in the Job Description.
\end{itemize}

The MasterTemplateFactory's output—the system message—is crucial for maintaining consistency and state across multiple LM interactions. It ensures that each LM query is grounded in the same comprehensive set of instructions, regardless of which section or subtask is being processed.

\subsubsection{Tools Implementation Module}
\label{subsubsec:tools_implementation_module}

The Tools Implementation module provides mechanisms for CELI to interact with external systems, execute specific functions, and manage data required for task completion across various domains.

\textbf{Key Responsibilities:}
\begin{itemize}
    \item Defining and managing tool functions that the LM can use.
    \item Supporting dynamic invocation of tools based on task requirements.
    \item Facilitating interactions with external APIs, databases, and computational resources.
\end{itemize}

\textbf{Core Features:}
\begin{itemize}
    \item Abstract base class \texttt{ToolImplementations} for defining tool functions.
    \item Dynamic generation of tool descriptions from method docstrings.
    \item Specialized implementations for diverse tasks (e.g., data retrieval, code execution, mathematical operations).
\end{itemize}

This module interfaces with the Processor and Section Processor modules, providing them with the necessary tools to execute complex tasks and interact with external resources. For example, it might include functions for retrieving mathematical formulas, executing Python code, or accessing specific datasets.

\subsubsection{Runner Module}
\label{subsubsec:runner_module}

The Runner module serves as the entry point for CELI execution, managing the configuration and initialization of the entire process.

\textbf{Key Responsibilities:}
\begin{itemize}
    \item Defining and managing the CELI configuration.
    \item Setting up the asynchronous execution environment.
    \item Initializing and executing the \texttt{ProcessRunner}.
    \item Managing LLM caching setup.
    \item Handling cleanup and resource management after execution.
\end{itemize}

\textbf{Core Components:}
\begin{itemize}
    \item \texttt{CELIConfig} class for centralizing configuration parameters (including system message).
    \item Asyncio event loop setup and management.
    \item \texttt{ProcessRunner} initialization and execution handler.
    \item LLM cache enabler and manager.
\end{itemize}

The Runner module acts as the orchestrator for the entire CELI process, from configuration to execution and cleanup. It provides a high-level interface for initiating CELI tasks, encapsulating the complexity of setup and resource management.

\subsubsection{Processor Module}
\label{subsubsec:processor_module}

The Processor module, implemented as the \texttt{ProcessRunner} class, manages the high-level execution flow of CELI tasks.

\textbf{Key Responsibilities:}
\begin{itemize}
    \item Initializing \texttt{SectionProcessor} instances for each section of the task.
    \item Managing the parallel or sequential execution of \texttt{SectionProcessors}.
    \item Handling the overall completion of the task.
\end{itemize}

\textbf{Core Components:}
\begin{itemize}
    \item \texttt{SectionProcessor} manager.
    \item Asynchronous execution handler.
    \item Completion notification system.
\end{itemize}

The Processor module manages the overall process flow, delegating the actual execution of tasks to individual \texttt{SectionProcessors}.

\subsubsection{Section Processor Module}
\label{subsubsec:section_processor_module}

The Section Processor module handles the execution of tasks within a specific section or subtask of the overall workflow, managing interactions with the LM and processing results.

\textbf{Key Features:}
\begin{itemize}
    \item Managing the iterative LLM interaction process for a specific section.
    \item Handling LLM responses and tool calls.
    \item Maintaining context for the section.
    \item Implementing error detection and recovery mechanisms.
    \item Conducting built-in reviews of section completion.
\end{itemize}

\textbf{Core Components:}
\begin{itemize}
    \item LLM interaction manager.
    \item Tool call dispatcher and executor.
    \item Context maintenance system.
    \item Error and loop detection mechanisms.
    \item Built-in review system.
\end{itemize}

The \texttt{SectionProcessor} is the core component that interacts directly with the LM, manages the execution flow within a section, and handles tool calls and error recovery. It operates independently for each section, guided by the initial system message and task specifications.

\subsection{Workflow Execution}
\label{subsec:workflow_execution}

CELI's workflow execution is a sophisticated process that combines the power of LMs with structured task management. The execution follows these key steps:

\begin{enumerate}
    \item \textbf{Task Initialization}: The \texttt{ProcessRunner} initializes the task based on the Job Description, creating \texttt{SectionProcessor} instances for each defined section.
    
    \item \textbf{Parallel or Sequential Execution}: Depending on the configuration, \texttt{SectionProcessors} are executed either concurrently using asyncio tasks or sequentially.
    
    \item \textbf{LM-Driven Execution}: Each \texttt{SectionProcessor} interacts with the LM, which controls the execution flow for its specific section. The LM receives the system message (created by the MasterTemplateFactory) and the current context, then decides on the next action.
    
    \item \textbf{Dynamic Tool Utilization}: As the LM processes the task, it may call for specific tools. These calls are handled by the \texttt{SectionProcessor}, which uses the \texttt{ToolImplementations} to execute the requested functions.
    
    \item \textbf{Context Management}: Throughout the execution, the \texttt{SectionProcessor} maintains and updates the context, ensuring that the LM has access to all relevant information for decision-making.
    
    \item \textbf{Error Handling and Recovery}: The system includes mechanisms for detecting errors or undesirable states (such as infinite loops). When such situations are detected, the \texttt{SectionProcessor} can initiate recovery procedures, including retrying the section or seeking human intervention if configured to do so.
    
    \item \textbf{Review and Refinement}: Built-in review mechanisms allow the system to evaluate the output of each section. If the output doesn't meet specified criteria, the section can be reprocessed with updated instructions or context.
    
    \item \textbf{Task Completion}: The process for each section continues until either a loop is detected (indicating a potential issue) or the LM explicitly calls the \texttt{complete\_section} tool, signaling that the section's objectives have been met.
    
    \item \textbf{Overall Task Finalization}: Once all sections have been processed, the \texttt{ProcessRunner} consolidates the results and performs any final cleanup or reporting tasks as specified in the Job Description.
\end{enumerate}

This execution flow allows CELI to handle complex, multi-stage tasks with a high degree of flexibility and robustness. The framework's ability to dynamically adapt to task requirements, handle errors, and maintain context throughout the process sets it apart from more rigid automation approaches.

\subsection{Advanced Features}
\label{subsec:advanced_features}

\subsubsection{Monitor Module}
\label{subsubsec:monitor_module}

The Monitor module is an experimental component of the CELI framework, designed to observe, evaluate, and potentially influence the performance and output of the \texttt{ProcessRunner}. It utilizes real-time data analysis to provide insights and maintain quality in automated tasks.

\textbf{Key Responsibilities:}
\begin{itemize}
    \item Observing and analyzing the \texttt{ProcessRunner}'s progress through log file monitoring.
    \item Processing queue messages to respond to specific events or states during operation.
    \item Utilizing AI models to assess task completion quality and make informed decisions.
    \item Providing feedback and recommendations for process improvements.
    \item Managing and tracking API token usage for efficient resource utilization.
\end{itemize}

\textbf{Core Components:}
\begin{itemize}
    \item \texttt{MonitoringAgent} class for overseeing \texttt{ProcessRunner} operations.
    \item Queue processing system for handling real-time updates and commands.
    \item AI-powered analysis system for evaluating prompt completions and function calls.
    \item Logging mechanism for tracking and reviewing monitoring activities.
    \item Token counter for managing API usage across monitoring operations.
\end{itemize}

The Monitor module ensures the quality and efficiency of the CELI framework's automated processes. By continuously analyzing outputs, detecting issues, and providing insights, it enables the system to adapt and improve over time, contributing to more reliable and effective document drafting and analysis tasks.

\subsubsection{Embeddor Module}
\label{subsubsec:embeddor_module}

The Embeddor module is an experimental component of the CELI framework, designed to transform pre-cleaned text data into embedded vectors suitable for machine learning applications.

\textbf{Key Responsibilities:}
\begin{itemize}
    \item Processing pre-cleaned text data from the Pre-processor module.
    \item Summarizing extensive text sections for efficient embedding.
    \item Filtering and chunking text data based on predefined schemas.
    \item Creating dataframes with embedded text data for analytical tasks.
    \item Managing token counts to optimize embedding processes.
\end{itemize}

\textbf{Core Components:}
\begin{itemize}
    \item Text summarization engine.
    \item Dictionary filtering system based on schema.
    \item Text chunking mechanism.
    \item Embedding generator using OpenAI's embeddings.
    \item Dataframe creator and manager.
\end{itemize}

The Embeddor module serves as a bridge between raw text data and machine learning models, transforming cleaned and standardized text into vector representations. This capability enhances CELI's ability to perform advanced analysis and processing tasks that require numerical representations of textual data.

\subsubsection{Mapper Module}
\label{subsubsec:mapper_module}

The Mapper module is an experimental component of the CELI framework, designed to manage the retrieval and processing of source materials for document drafting tasks. It focuses on mapping and identifying essential sections from source documents to support automated document analysis.

\textbf{Key Responsibilities:}
\begin{itemize}
    \item Mapping document sections to corresponding source materials.
    \item Pre-computing essential sources for document sections.
    \item Managing retries for failed section retrievals.
    \item Saving pre-computed essential source mappings and error lists.
    \item Facilitating efficient processing of document data for drafting and analysis.
\end{itemize}

\textbf{Core Components:}
\begin{itemize}
    \item \texttt{EssentialSourceFacade} class for managing source material retrieval and analysis.
    \item Mapping system for document sections to source materials.
    \item Checkpoint mechanism for saving and loading progress.
    \item Error detection and handling system for processing results.
    \item Caching system for efficient retrieval of previously processed data.
\end{itemize}

The Mapper module serves as a crucial link between source documents and the drafting process, ensuring that relevant information is efficiently identified and processed. It works in conjunction with other CELI modules to enhance the accuracy and completeness of document drafts by providing structured access to essential source materials.

\subsection{Conclusion}
\label{subsec:conclusion}

The CELI framework represents a significant advancement in integrating Language Models into complex computational systems. By embedding control logic directly within LM prompts and providing a modular, flexible architecture, CELI enables the execution of sophisticated, multi-stage tasks across diverse domains. The core components and advanced features described in this appendix work in concert to create a powerful system capable of adapting to a wide range of applications, from data analysis and report generation to content creation and automated evaluations. As CELI continues to evolve, its components and features are being actively refined and expanded, enhancing the framework's capabilities and broadening its applicability across various fields.

\lstset{
  breaklines=true,
  breakatwhitespace=true,
  basicstyle=\ttfamily\small,
  columns=flexible,
  keepspaces=true
}

\section{HumanEval Case Study Details}
\label{sec:appendix_b}

\subsection{Job Description Tasks (Reported Verbatim)}
\label{subsection:humaneval_job_description}

\subsubsection{Task 1: Retrieve Coding Prompt}
\begin{verbatim}
Task(
  task_name="Get the prompt",
  details={
    "description": "Find the coding prompt for the current section by calling get_prompt",
  },
)
\end{verbatim}

\subsubsection{Task 2: Develop Test Cases}
\begin{lstlisting}[breaklines=true, basicstyle=\small\ttfamily]
Task(
  task_name="Develop initial test cases",
  details={
    "description": """Think through the problem and develop an initial set of test cases that will be used to check the logic for your function. The test cases must be implemented as a function called 'def check(candidate)' that takes the function as an argument. Each test is an 'assert' statement that calls the function with some inputs and verifies that that output is as expected. For example, if the task is to create a function called 'add_two_numbers", the test function might look like:
    def check(candidate):
        assert add_two_numbers(1, 2) == 3
        assert add_two_numbers(0, 0) == 0
        assert add_two_numbers(-1, 1) == 0
        assert add_two_numbers(1.5, 1) == 2.5
    
    When writing test cases, think through edge cases and different types of inputs that might be passed to the function. Some functions that may seem very simple have a trick to them. When mapping from a real world description of a problem to the algorithm, make sure you have the situation modeled correctly. Be sure the function behaves correctly when numbers are integers or floats. Be sure to handle negative numbers appropriately. If you have a test case that is failing, write out the intermediate steps involved in that test case and see if that explains why your implementation is failing. If you need to debug, you can add print statements into the functions you pass to run_tests. Think about issues like empty inputs and multiple delimiters.
    
    Remember that check must take the 'candidate' argument that is the function to test.""",
  },
)
\end{lstlisting}

\subsubsection{Task 3: Implement and Test Function}
\begin{lstlisting}
Task(
  task_name="Write and test code",
  details={
    "description": """In this task you iteratively refine your implementation and test cases. Decide how to implement the function and call run_tests to check your implementation, passing in the code and tests. Make sure to include all required imports. If the tests don't pass, check both your implementation and the test cases and decide which needs to be adjusted (or both). Don't assume the test cases are correct, review the problem specification and adjust if necessary. Also, feel free to add more tests. Keep going as long as you are making progress, but if you can't get all the tests to pass after a few tries, just save your output and complete.""",
  },
)
\end{lstlisting}

\subsubsection{Task 4: Finalize and Save Output}
\begin{lstlisting}
Task(
  task_name="Produce the final output.",
  details={
    "description": """Call save_final_output to save off your final answer. This should include your function implementation as well as your test function. Always do this, even if you aren't completely happy with your answer. Signal that you have completed the example by calling the complete_section function."""
  },
)
\end{lstlisting}

\subsection{Tool Implementations} \label{subsection:humnaneval_tool_implementations}

\subsubsection{Tool 1: Retrieve Coding Prompt}
\noindent\textbf{Function \texttt{get\_prompt(task\_id: str) -> str}:}
\begin{itemize}
    \item \textbf{Description}: Returns the prompt for the given task.
    \item \textbf{Args}:
    \begin{itemize}
        \item \texttt{task\_id} (str): The task id, including the "HumanEval/" prefix. For example, "HumanEval/10".
    \end{itemize}
    \item \textbf{Returns}: str - The prompt for this question. This will be a function signature and doc string.
    \item \textbf{Implementation}:
    \begin{lstlisting}[language=Python]
return self.tests.loc[task_id, "prompt"]
    \end{lstlisting}
\end{itemize}

\subsubsection{Tool 2: Execute Tests in Sandbox}
\noindent\textbf{Function \texttt{run\_tests(func, test\_func)}:}
\begin{itemize}
    \item \textbf{Process}:
    \begin{itemize}
        \item \texttt{sandbox} $\leftarrow$ \texttt{create\_sandboxed\_environment()}
        \item \texttt{result} $\leftarrow$ \texttt{sandbox.execute(func, test\_func)}
        \item Return \texttt{result} with failing assertions to context
    \end{itemize}
\end{itemize}

\subsubsection{Tool 3: Save Final Output}
\noindent\textbf{Function \texttt{save\_final\_output(task\_id: str, func: str, test\_func: str)}:}
\begin{itemize}
    \item \textbf{Description}: Writes the final output for the problem.
    \item \textbf{Args}:
    \begin{itemize}
        \item \texttt{task\_id} (str): The HumanEval problem identifier
        \item \texttt{func} (str): The body of the function indented by 4 spaces (without the function signature).
        \item \texttt{test\_func} (str): The full definition of the \texttt{check} function (including the \texttt{def check} line)
    \end{itemize}
    \item \textbf{Note}: This must be called before calling \texttt{complete\_section} and moving on to the next problem. The most recently submitted code will be used.
    \item \textbf{Implementation}: Logs the completed problem, saves the code and tests to a JSON file.
\end{itemize}

\subsection{Example Log} \label{subsection:example_log}

Task execution for \texttt{HumanEval/2} is paraphrased from the logs as follows:

\begin{itemize}
    \item \textbf{LM MESSAGE}: "Let's think through the problem and develop test cases."
    \item \textbf{TOOL CALL}: \texttt{GET\_PROMPT("HumanEval/2")}
    \item \textbf{RESULT}: Received function signature for \texttt{truncate\_number(number: float) -> float}
    \item \textbf{LM MESSAGE}: "The function should handle:
    \begin{itemize}
        \item Numbers with non-zero decimal parts
        \item Numbers with zero decimal parts
        \item Very small decimal parts
        \item Large numbers with decimal parts
        \item Numbers very close to integers"
    \end{itemize}
    \item \textbf{LM MESSAGE}: "Let's create the \texttt{check(candidate)} function with these test cases:"
    \item \textbf{LM MESSAGE}:
    \begin{lstlisting}[language=Python]
def check(candidate):
    # Test cases
    assert candidate(3.5) == 0.5
    assert candidate(4.0) == 0.0
    assert candidate(0.123) == 0.123
    assert candidate(123456.789) == 0.789
    assert candidate(1.999999) == 0.999999
    \end{lstlisting}
    \item \textbf{LM MESSAGE}: "Now that we have our test cases, let's implement the function and run the tests."
    \item \textbf{LM MESSAGE}: "Let's implement the function and run the tests."
    \item \textbf{LM MESSAGE}:
    \begin{lstlisting}[language=Python]
def truncate_number(number: float) -> float:
    return number - int(number)
    \end{lstlisting}
    \item \textbf{TOOL CALL}: \texttt{RUN\_TESTS}
    \item \textbf{RESULT}: \textit{ERROR: Missing 'test\_func' argument in tool call} (from logged Exception)
    \item \textbf{LM MESSAGE}: "I missed providing the \texttt{test\_func} argument. Let's correct that and run again."
    \item \textbf{TOOL CALL}: \texttt{RUN\_TESTS} (Corrected)
    \item \textbf{RESULT}: \textit{FAILED: assert candidate(123456.789) == 0.789}
    \item \textbf{LM MESSAGE}: "This might be due to floating-point precision issues. Let's update the function."
    \item \textbf{LM MESSAGE}: "We'll use the \texttt{round} function to ensure accuracy."
    \item \textbf{LM MESSAGE}:
    \begin{lstlisting}[language=Python]
def truncate_number(number: float) -> float:
    return round(number - int(number), 10)
    \end{lstlisting}
    \item \textbf{TOOL CALL}: \texttt{RUN\_TESTS}
    \item \textbf{RESULT}: All tests \textit{PASSED}
    \item \textbf{LM MESSAGE}: "The function has passed all test cases. Let's save the final implementation."
    \item \textbf{TOOL CALL}: \texttt{SAVE\_FINAL\_OUTPUT("HumanEval/2", func, test\_func)}
    \item \textbf{LM MESSAGE}: "Now that we've saved the output, let's complete the section."
    \item \textbf{TOOL CALL}: \texttt{COMPLETE\_SECTION()}
    \item \textbf{RESULT}: Successfully implemented and tested \texttt{truncate\_number} function
\end{itemize}

\subsection{Selection of Evaluation Tools} \label{subsection:Selection_of_Evaluation}

CELI's implementation for HumanEval utilizes GPT-4-Turbo (0125) \citep{openai2024gpt4}, which was chosen after careful consideration and testing. Notably, many of the specialized coding agents we aimed to compare against used GPT-4-Turbo (1106). However, in our experiments with CELI, we found that the 1106 version did not consistently follow instructions or reliably execute tool calls to the degree required for our framework's approach. In contrast, GPT-4-Turbo (0125), despite being an older version, demonstrated superior adherence to instructions and more reliable tool call execution within CELI's architecture.

\section{Wikipedia Article Case Study Details}
\label{sec:Case_Study_Details}

\subsection{Job Description Tasks (Reported Verbatim)}
\label{subsection:job_description}

\subsubsection{Task 1: Search for Example Document Section Text}
\begin{lstlisting}
Task(
  task_name="Search for Document Section Text",
  details={
    "description": "Find the text of the specified section and all subsections in the example document.",
    "prerequisite_tasks": [],
    "function_call": "Call get_example_toc to get the full list of sections in the example doc and then get_text_for_sections to retrieve the text for the specified section and any relevant (sub)sections.",
    "example_call": "{{'Example Document': ['1', '1.1', '1.2']}}",
    "instructions": [
      "Use get_example_and_target_names to get the name of the target that the document should be about.",
      "The specified section should have corresponding text, even that text is blank. If you get an error, try again with different parameters",
      "Do not truncate or modify the retrieved text.",
      "If text is present, print the entire text and instruct to proceed to the next task.",
      "If the specified section is empty in the example document, then you can leave it empty in the target. In that case, you can skip all the remaining tasks and jump straight to the 'Draft New Document Section' task, which can just draft an empty section. Note that a blank section is not the same as the function returning an error.",
    ],
  },
)
\end{lstlisting}

\subsubsection{Task 2: Understand Differentiation}
\begin{lstlisting}
Task(
  task_name="Understand Differentiation",
  details={
    "description": "Understand the context of the example document section by comparing it with similar sections. Also look at any subsections and understand how they are structured.",
    "prerequisite_tasks": ["Search for Document Section Text"],
    "instructions": [
      "Identify other sections of the example document that may contain content similar to the current section.",
      "Retrieve the text of these sections along with their section identifiers.",
      "Analyze and note down how the current section differs from these sections to prevent duplication in future work.",
    ],
    "additional_notes": [
      "Keep your notes concise and relevant for later use.",
      "If 'no content present' is observed in all section bodies that are retrieved, even after retrieving children/sub-sections, proceed to the next task.",
      "If it seems like the current section is specific to the example document, and would not make sense as part of the target document, feel free to skip the section and go on to the next one. This can happen if the example contains a section highly specific to its topic, but not relevant to the target document.",
    ],
  },
)
\end{lstlisting}

\subsubsection{Task 3: Find Relevant References}
\begin{lstlisting}
Task(
  task_name="Find the most relevant references for the target section and all the subsections",
  details={
    "description": """Call get_corresponding_target_references_for_example_sections function with a list of the current section and all subsections to retrieve these materials. to find relevant references for the target document that correspond to the example document sections you are looking at.
    The references may not provide you with all the information you need to draft the section. Don't worry, you will get a chance to ask additional questions in the next task.""",
    "example_call": "['1', '1.1', '1.2']",
  },
)
\end{lstlisting}

\subsubsection{Task 4: Ask Additional Questions}
\begin{lstlisting}
Task(
  task_name="Ask additional questions",
  details={
    "description": """Determine if any additional information is required to draft the section and call the 
    ask_question_about_target function to gather that information. Remember, your goal is to create a page about
    the target, not the example. Feel free to ask as many questions as you need and keep working on this task 
    until you have the information you need. This is especially important if the initial references turn out 
    not to be useful. If you ask questions, make sure they are specific and ask directly about the target by 
    name. Do not ask questions about the example. For 
    example, ask "What are major events in Henry Thoreau's life" instead of "What are major events in the author's life?"
    If you don't have any questions, just move on to the next section.""",
    "example_call": '{"prompt": "What is unusual about the formation of Cream?"}',
  },
)
\end{lstlisting}

\subsubsection{Task 5: Define Subsections for this Section}
\begin{lstlisting}
Task(
  task_name="Define subsections for this section",
  details={
    "description": """Define what subsections should be present within this individual section. Use the table 
    of contents from the example document and your knowledge of the target to structure the subsections.  
    Keep in mind over differentiation of this section from other sections in the document. It is totally fine 
    to not have subsections, especially if the example document does not have them.
    Also, remember that the subsections should be relevant for the target document. The detailed structure of 
    subsections used in the example may not be relevant for our target document.""",
  },
)
\end{lstlisting}

\subsubsection{Task 6: Draft New Document Section}
\begin{lstlisting}
Task(
  task_name="Draft New Document Section",
  details={
    "description": "Draft a new section analogous to the example section, but about the target subject. Ensure alignment with its "
    "structure, format, and scope (from {{TaskRef:Understand Differentiation}} output). Use "
    "the section structure you defined in {{TaskRef:Define subsections for this section}}. "
    "However, the details should be related to the target and not the example document."
    "Call the save_draft_section tool to save the draft. Pass in the section number to save_draft_section",
    "guidelines": [
      "Clearly identify the section number and section heading/title at the top of the content.",
      "The new section should have its unique scope and purpose, distinct from the example section.",
      "Avoid duplicating content or including redundant information.",
      "Aim for the new section to mirror the example section in length and detail, but using content related to the target.",
      "Follow the instructions set out by {{TaskRef:Understand Differentiation}} output.",
      "Maintain consistency in documentation methodology, using the revised example as a template.",
      "Ensure content is exclusively about the target, and not the example topic.",
    ],
    "specific_instructions": [
      "Do not copy text verbatim. Include only text within the scope of the current section, as highlighted in the output of {{TaskRef:Understand Differentiation}}.",
      "Include cross-references to other sections as seen in the example if applicable.",
    ],
  },
)
\end{lstlisting}

\subsubsection{Task 7: Prepare for Next Document Section}
\begin{lstlisting}
Task(
  task_name="Prepare for Next Document Section",
  details={
    "description": "Signal that you have completed the draft by calling complete_section",
    "function_call": "Use the complete_section function with the argument value = current section identifier.",
    "example_call": "{{'current_section_identifier': ['1.2']}}",
  },
)
\end{lstlisting}

\subsection{Tool Implementations} \label{subsection:tool_implementations}

\subsubsection{Tool 1: Retrieve Example and Target Names}
\noindent\textbf{Function \texttt{get\_example\_and\_target\_names() -> Dict[str, str]}:}
\begin{itemize}
    \item \textbf{Description}: Gets the names for the example and target documents.
    \item \textbf{Returns}: A dictionary with keys 'example' and 'target', containing the respective document names.
    \item \textbf{Implementation}:
    \begin{lstlisting}[language=Python]
return {
    "example": self.example_page,
    "target": self.target_page,
}
    \end{lstlisting}
\end{itemize}

\subsubsection{Tool 2: Retrieve Example Table of Contents}
\noindent\textbf{Function \texttt{get\_example\_toc() -> str}:}
\begin{itemize}
    \item \textbf{Description}: Retrieves and formats the table of contents for the example document.
    \item \textbf{Returns}: A string containing the formatted table of contents for the example document.
    \item \textbf{Implementation}:
    \begin{lstlisting}[language=Python]
return f"Table of Contents:\n\n{format_toc(self.schema)}"
    \end{lstlisting}
\end{itemize}

\subsubsection{Tool 3: Retrieve Text for Sections}
\noindent\textbf{Function \texttt{get\_text\_for\_sections(sections\_dict\_str: str) -> str}:}
\begin{itemize}
    \item \textbf{Description}: Extracts text from specified sections of documents.
    \item \textbf{Args}:
    \begin{itemize}
        \item \texttt{sections\_dict\_str} (str): A JSON string mapping document names to their respective section numbers.
    \end{itemize}
    \item \textbf{Returns}: Concatenated text from the specified sections or error message if there was a problem.
    \item \textbf{Process}:
    \begin{itemize}
        \item Parse \texttt{sections\_dict\_str} into Python dictionary
        \item For each document and its sections:
        \begin{itemize}
            \item Load document content
            \item Extract text for specified sections
            \item Concatenate extracted text
        \end{itemize}
        \item Return concatenated text or error message
    \end{itemize}
\end{itemize}

\subsubsection{Tool 4: Retrieve Corresponding Target References}
\noindent\textbf{Function \texttt{get\_corresponding\_target\_references\_for\_example\_sections(section\_numbers: List[str]) -> str}:}
\begin{itemize}
    \item \textbf{Description}: Retrieves target document references corresponding to example document sections.
    \item \textbf{Args}:
    \begin{itemize}
        \item \texttt{section\_numbers} (List[str]): A list of section numbers from the example document.
    \end{itemize}
    \item \textbf{Returns}: A string containing the most relevant reference content for the target document.
    \item \textbf{Process}:
    \begin{itemize}
        \item reference\_ids $\leftarrow$ \_get\_references\_for\_example\_sections(section\_numbers)
        \item all\_nodes $\leftarrow$ []
        \item For each ref in reference\_ids:
        \begin{itemize}
            \item If using ChromaVectorStore:
            \begin{itemize}
                \item query\_embedding $\leftarrow$ example\_index.vector\_store.get\_embedding(ref)
            \end{itemize}
            \item Else:
            \begin{itemize}
                \item query $\leftarrow$ example\_index.docstore.get\_node\_content(ref)
            \end{itemize}
            \item related\_nodes $\leftarrow$ target\_retriever.retrieve(query or query\_embedding)
            \item all\_nodes.extend(related\_nodes)
        \end{itemize}
        \item top\_nodes $\leftarrow$ \_get\_top\_nodes(all\_nodes, 10)
        \item Return concatenated and normalized content of top\_nodes
    \end{itemize}
\end{itemize}

\subsubsection{Tool 5: Ask Question About Target}
\noindent\textbf{Function \texttt{ask\_question\_about\_target(prompt: str) -> str}:}
\begin{itemize}
    \item \textbf{Description}: Asks a question about the target subject and returns a response using a vector database query engine.
    \item \textbf{Args}:
    \begin{itemize}
        \item \texttt{prompt} (str): The question to ask about the target subject.
    \end{itemize}
    \item \textbf{Returns}: The response to the question about the target subject.
    \item \textbf{Process}:
    \begin{itemize}
        \item Cache $\leftarrow$ get\_celi\_llm\_cache()
        \item result $\leftarrow$ cache.check\_llm\_cache(target\_url, prompt)
        \item If result exists in cache:
        \begin{itemize}
            \item Return \_dict\_to\_response(result["response"])
        \end{itemize}
        \item Else:
        \begin{itemize}
            \item result $\leftarrow$ target\_query\_engine.query(prompt)
            \item result\_dict $\leftarrow$ \_response\_to\_dict(result)
            \item Cache result\_dict
            \item Return result.response
        \end{itemize}
    \end{itemize}
    \item \textbf{Note}: This method utilizes a query engine (\texttt{target\_query\_engine}) initialized with the target document's index in a vector database. The query engine processes the prompt and retrieves relevant information from the vector store to generate the response.
\end{itemize}

\subsubsection{Tool 6: Save Draft Section}
\noindent\textbf{Function \texttt{save\_draft\_section(section\_number: str, content: str)}:}
\begin{itemize}
    \item \textbf{Description}: Saves the draft content for a specific section.
    \item \textbf{Args}:
    \begin{itemize}
        \item \texttt{section\_number} (str): The section number to save the draft for.
        \item \texttt{content} (str): The content of the draft section.
    \end{itemize}
    \item \textbf{Implementation}: Saves the draft content to a file or database.
\end{itemize}

\subsection{Wikipedia Article Scoring and Evaluation Prompt} \label{subsection:Evaluation_Prompt}

\textit{Your job is to compare a generated document versus a human-created reference. Rate the document using an integer from 0-6 using the scale below.}

\begin{itemize}

    \item \textit{0 - Irrelevant: The AI document is completely off-topic or unusable.}
    \item \textit{1 - Very Poor: Major errors or missing information make the document largely ineffective.}
    \item \textit{2 - Insufficient: Significant elements are missing, and extensive revisions are needed.}
    \item \textit{3 - Marginal: Meets the basic requirements but contains several deficiencies.}
    \item \textit{4 - Satisfactory: Acceptable as a first draft but requires refinement.}
    \item \textit{5 - Comparable: Matches the quality and completeness of the ground truth document.}
    \item \textit{6 - Outstanding: Surpasses the ground truth in quality, detail, and presentation.\\Have the scores be discrete (no floats).}
\end{itemize}

\subsection{Example Log} \label{subsection:Example_Log}

Task execution for the \textit{Jonas Brothers} article generation is paraphrased from the logs as follows:

\begin{itemize}
    \item \textbf{LM MESSAGE}: "Let's start by retrieving the example document's structure and content."
    \item \textbf{TOOL CALL}: \texttt{GET\_EXAMPLE\_TOC()}
    \item \textbf{RESULT}: Received table of contents for Led Zeppelin article
    \item \textbf{LM MESSAGE}: "Now, let's get the text for the main section of the Led Zeppelin article."
    \item \textbf{TOOL CALL}: \texttt{GET\_TEXT\_FOR\_SECTIONS(\{"Led Zeppelin": ["0"]\})}
    \item \textbf{RESULT}: Retrieved text for the main section of Led Zeppelin article
    \item \textbf{LM MESSAGE}: "Analyzing the structure and content of the Led Zeppelin article to understand how to approach the Jonas Brothers article."
    \item \textbf{LM MESSAGE}: "Let's find relevant references for the Jonas Brothers article."
    \item \textbf{TOOL CALL}: \texttt{GET\_CORRESPONDING\_TARGET\_REFERENCES\_FOR\_EXAMPLE\_SECTIONS(["0"])}
    \item \textbf{RESULT}: Retrieved relevant references for Jonas Brothers
    \item \textbf{LM MESSAGE}: "Now that we have some references, let's ask additional questions to fill in any gaps."
    \item \textbf{LM MESSAGE}: "Questions to Ask:
    \begin{enumerate}
        \item What are the major milestones in the career of the Jonas Brothers?
        \item How did the Jonas Brothers form and what were their early years like?
        \item What are some of the most significant achievements and awards received by the Jonas Brothers?
        \item How has the musical style of the Jonas Brothers evolved over the years?
        \item What is the legacy and impact of the Jonas Brothers on the music industry?
    \end{enumerate}"
    \item \textbf{TOOL CALL}: \texttt{ASK\_QUESTION\_ABOUT\_TARGET("What are major milestones in the Jonas Brothers' career?")}
    \item \textbf{RESULT}: Received information about Jonas Brothers' career milestones
    \item \textbf{LM MESSAGE}: "Let's define the subsections for the Jonas Brothers article based on our analysis and the information we've gathered."
    \item \textbf{LM MESSAGE}: "Proposed Subsections for Section 0 (Jonas Brothers):
    \begin{enumerate}
        \item Introduction
        \item Formation and Early Years
        \item Major Milestones
        \item Musical Style and Evolution
        \item Achievements and Awards
        \item Legacy and Impact
    \end{enumerate}"
    \item \textbf{LM MESSAGE}: "Now, let's draft the main section of the Jonas Brothers article."
    \item \textbf{TOOL CALL}: \texttt{SAVE\_DRAFT\_SECTION("0", [Content of drafted section])}
    \item \textbf{RESULT}: Draft saved successfully
    \item \textbf{LM MESSAGE}: "Task Completed: Task 6 (Draft New Document Section)"
    \item \textbf{TOOL CALL}: \texttt{COMPLETE\_SECTION()}
    \item \textbf{RESULT}: Successfully completed article generation for Jonas Brothers
\end{itemize}

\pagebreak
\subsection{Example Evaluations} \label{subsection:Example_Evaluations}

\begin{table}[h]
\centering
\caption{Evaluations of generated wiki articles by Claude 3.5 Sonnet.}
\label{tab:wiki_evaluations}
\resizebox{\textwidth}{!}{%
\begin{tabular}{|c|c|c|c|p{7.5cm}|}
\hline
\textbf{Category} & \textbf{Example} & \textbf{Target} & \textbf{Score} & \textbf{Key Observations} \\
\hline
Bands & Led Zeppelin & Jonas Brothers & 4 & Structured overview, some inaccuracies, lacks cultural impact depth \\
\hline
Bands & Led Zeppelin & Daft Punk & 1 & Numerous inaccuracies, wrong band details, disjointed structure \\
\hline
Bands & Jonas Brothers & Led Zeppelin & 4 & Comprehensive, lacks cultural impact and historical details \\
\hline
Bands & Jonas Brothers & Daft Punk & 4 & Broad overview, lacks early career details, some formatting issues \\
\hline
Bands & Daft Punk & Led Zeppelin & 5 & Well-organized, covers key aspects, lacks some specific details \\
\hline
Bands & Daft Punk & Jonas Brothers & 4 & Comprehensive, some discography inaccuracies, some repetition \\
\hline
Countries & Ghana & Jamaica & 4 & Comprehensive, well-structured, lacks deep context in some areas \\
\hline
Countries & Ghana & Belize & 0 & Completely irrelevant content, covers Ghana instead of Belize \\
\hline
Countries & Jamaica & Ghana & 3 & Detailed on Ghana, includes irrelevant Jamaica content \\
\hline
Countries & Jamaica & Belize & 3 & Comprehensive but misaligned, some inaccuracies and redundancies \\
\hline
Countries & Belize & Ghana & 5 & Well-organized, highly detailed, lacks engaging narrative flow \\
\hline
Countries & Belize & Jamaica & 4 & Thorough, some factual inaccuracies, lacks professional polish \\
\hline
Drugs & Sirolimus & Prednisone & 5 & Thorough and structured, lacks some nuanced details \\
\hline
Drugs & Sirolimus & Semaglutide & 4 & Comprehensive overview, lacks specific details and fluid narrative \\
\hline
Drugs & Prednisone & Sirolimus & 5 & Comprehensive coverage, some formatting inconsistencies \\
\hline
Drugs & Prednisone & Semaglutide & 4 & Well-organized, missing nuanced elements and broader context \\
\hline
Drugs & Semaglutide & Sirolimus & 4 & Detailed content, lacks depth in some areas, fewer citations \\
\hline
Drugs & Semaglutide & Prednisone & 5 & Comprehensive, well-structured, possible minor inaccuracies \\
\hline
\end{tabular}
}
\end{table}

\end{document}